\documentclass[aps,prl,floatfix,twocolumn]{revtex4}
\usepackage{graphics}
\usepackage{dcolumn}
\usepackage{epsfig}

\newcommand{\bk}{{\bf k }}

\newcommand{\br}{{\bf r }}

\newcommand{\bG}{{\bf G}}

\newcommand{\bK}{{\bf K}}

\begin{document}

\title{Electron Beam Supercollimation in Graphene Superlattices}

\author{Cheol-Hwan Park$^{1,2}$}
\author{Young-Woo Son$^{3,4}$}
\author{Li Yang$^{1,2}$}
\author{Marvin L. Cohen$^{1,2}$}
\author{Steven G. Louie$^{1,2}$}
\email{sglouie@berkeley.edu}
\affiliation{$^1$Department of Physics, University of California at Berkeley,
Berkeley, California 94720 USA\\
$^2$Materials Sciences Division, Lawrence Berkeley National
Laboratory, Berkeley, California 94720 USA\\
$^3$Department of Physics, Konkuk University, Seoul 143-701, Korea\\
$^4$School of computational sciences, Korea institute for advanced study,
Seoul 130-722, Korea}
\date{\today}

\begin{abstract}
Although electrons and photons are intrinsically different, importing
useful concepts in optics to electronics performing similar functions has been
actively pursued over the last two decades. In particular, collimation of an
electron beam is a long-standing goal.
We show that ballistic propagation
of an electron beam with virtual no spatial spreading or diffraction,
without a waveguide or external magnetic field, can be achieved
in graphene under an appropriate
class of experimentally feasible one-dimensional external periodic potentials.
The novel chiral quasi-one-dimensional metallic state that the charge carriers are in
originates from a collapse of the intrinsic
helical nature of the charge carriers in graphene owing to the superlattice potential.
Beyond providing a new way to constructing chiral one-dimensional states in two dimensions,
our findings should be useful in graphene-based electronic devices (e.\,g.\,, for information
processing) utilizing some of the highly developed concepts in optics.
\end{abstract}
\maketitle

Electronic analogues of many optical behaviors such as 
focusing~\cite{Houten:EPL,Spector:APL,Sivan:PRB}, 
collimation~\cite{Molenkamp:PRB}, 
and interference~\cite{Yacoby:PRL} have been achieved 
in two-dimensional electron gas (2DEG), enabling the system 
as a basic platform to study fundamental problems 
in quantum mechanics~\cite{Buks:Nat,Yang:Nat,Chang:NatPhys} 
as well as quantum information 
processing~\cite{Velsen:PRL}.
The close relationship between optics 
and electronics is been made possible 
due to the ballistic transport properties 
of a high-mobility 2DEG created  
in semiconductor heterostructures~\cite{Palevski:PRL}. 
Among those electronics-optics analogues, 
the collimation or quasi-one-dimensional (quasi-1D) motion 
of electrons and photons are particularly important 
not only to achieve electronic quantum 
devices~\cite{Yang:Nat,Velsen:PRL} 
but also to realize ultracompact integrated 
light circuits~\cite{kosaka:1212,wu:2003JLT}.
Usually, electrons originated from a point source 
may be controlled by electrostatics 
or geometrical constrictions~\cite{Spector:APL,Sivan:PRB}.
Quasi-1D electronic states and focusing 
have been achieved in a 2DEG with the help of external magnetic 
fields, e.\,g.\,, employing magnetic focusing~\cite{Houten:EPL}
and quantum Hall edge states~\cite{Yang:Nat,Prange:book}.
However, it would be difficult to integrate them into a single electronic 
device due to the external high magnetic field apparatus needed.
In view of recent successful demonstrations of extreme 
anisotropic light propagation without diffraction, 
called supercollimation in photonic 
crystals~\cite{kosaka:1212,wu:2003JLT,rakich:2006nmat,joannopoulus:book}, 
an analogue of this effect in two-dimensional (2D) electron systems 
may also be possible.
In this work, we demonstrate that graphene~\cite{novoselov:2005Nat_Graphene_QHE,
zhang:2005Nat_Graphene_QHE,berger:2006Graphene_epitaxial}
in an external periodic potentials,
or a graphene superlattice, is particularly suitable to realize 
electron supercollimation in two dimensions.

The isolation of graphene~\cite{novoselov:2005Nat_Graphene_QHE,
zhang:2005Nat_Graphene_QHE,berger:2006Graphene_epitaxial}, 
a single layer of carbon atoms 
in a honeycomb structure composed of two equivalent sublattices,
offers a new dimension to study electronics-optics analogues.
Carriers in graphene exhibit ballistic transport on the submicron scale 
at room temperature~\cite{Shedin:NatMat}
and with mobility up to
$2\times10^5\,$cm$^2\,$V$^{-1}\,$s$^{-1}$~\cite{Bolotin:SSC}.
Graphene electronic states have an internal quantum number, a pseudospin, that is not found 
in normal electronic systems and strongly influences the dynamics of the charge carriers.
The pseudospin is of central importance to many of the novel physical properties
of graphene~\cite{novoselov:2005Nat_Graphene_QHE,
zhang:2005Nat_Graphene_QHE,berger:2006Graphene_epitaxial,
Shedin:NatMat,Bolotin:SSC,katsnelson:2006NatPhys_Graphene_Klein,Altshuler:Sci},
and it also plays a significant role in the present work.

\begin{figure*}
\includegraphics[width=1.8\columnwidth]{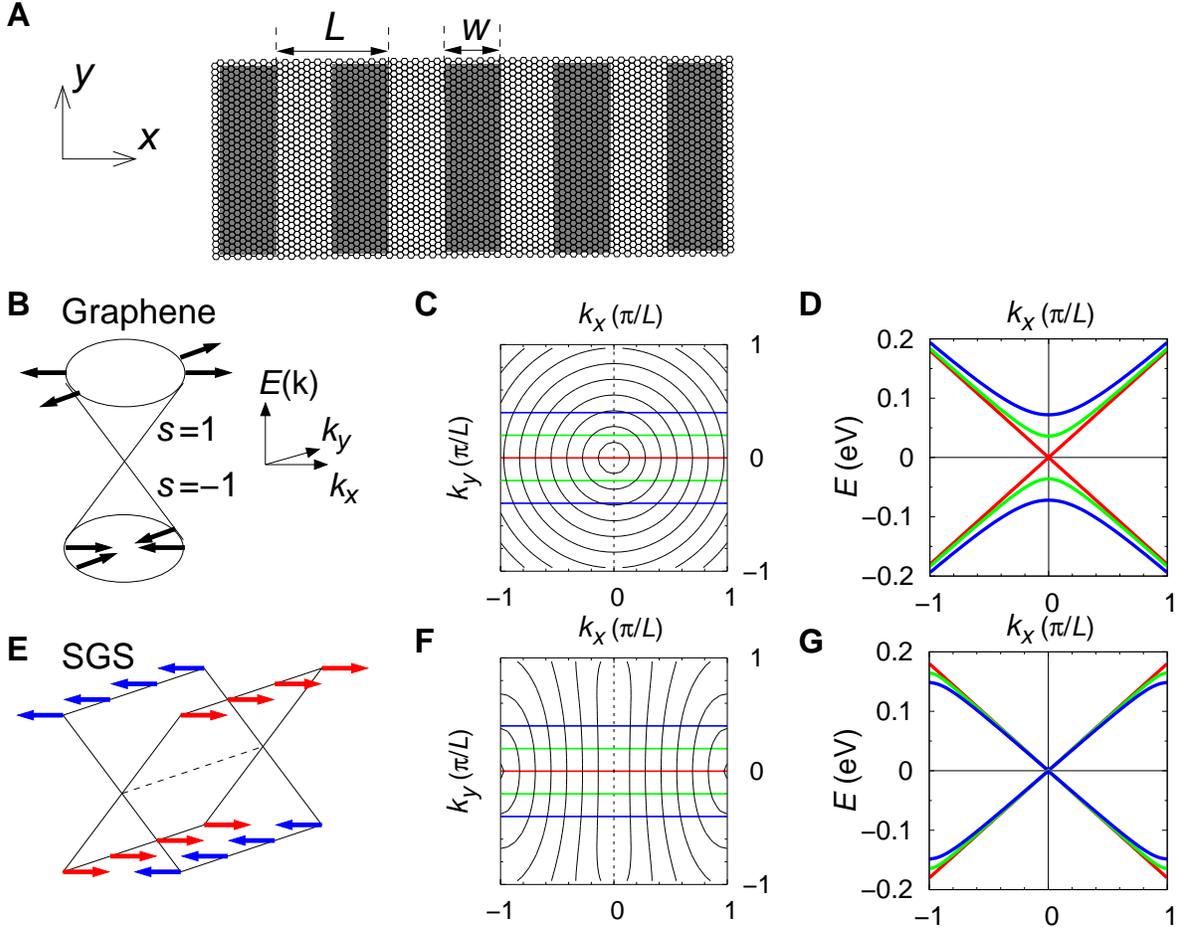}
\caption{{Electron energy dispersion relation in a special graphene superlattice.}
({\bf A}) Schematic diagram of a Kronnig-Penney type of potential
applied to graphene with strength $U_0$ inside the gray regions and zero outside.
The lattice period is $L$ and the barrier width is $w$.
({\bf B}) Schematic diagram showing the electronic energy dispersion relations
and pseudospin vectors (black arrows) in graphene.
({\bf C}) Contour plot of the first electronic band above the Dirac point energy
in pristine graphene. The energy difference between neighbouring contours is 25~meV,
with the lowest contour near the origin having a value of 25~meV.
({\bf D}) The electronic energy dispersion relation $E$ versus $k_x$ with fixed $k_y$.
Red, green and blue lines correspond to $k_y=0$, $0.1\ \pi/L$ and
$0.2\ \pi/L$, respectively, as indicated in {\bf C}.
({\bf E}), ({\bf F}) and ({\bf G})
Same quantities as in {\bf B}, {\bf C} and {\bf D} for the considered
SGS ($U_0=0.72$~eV, $L=10$~nm and $w=$~5nm).
Red and blue arrows in {\bf E} represent the `right' and the `left'
pseudospin state, respectively.}
\label{Fig1}
\end{figure*}

The low-energy quasiparticles in graphene
whose wavevectors are close to
the Dirac point $\bK$ in the Brillouin zone
are described by a $2\times2$ Hamiltonian matrix
$H_0(\bk)=\hbar v_0 \left(\sigma_xk_x+\sigma_yk_y\right)$,
where $v_0\approx10^6$~m/s is the band velocity,
$\bk$ is the
wavevector measured from the $\bK$ point, and
$\sigma$'s are the Pauli matrices.
The energy eigenvalues are given by 
$
E^0_{s}(\bk)=s\,\hbar\, v_0\, |{\bf k}|
$
where $s=+1\,(-1)$ denotes the conical conduction (valence) band (Fig.~1B).
The sublattice degree of freedom of the quasiparticles in graphene
can conveniently be described with a pseudospin basis, or spinors, where the
$\left|\uparrow\right>$ and the $\left|\downarrow\right>$ pseudospin states
of $\sigma_z$ represent $\pi$-electron orbital on the {\it A} and {\it B} 
sublattices of the structure of graphene, respectively.
This Hamiltonian is very similar to the one used to model neutrinos
as massless Dirac fermions~\cite{Ando:JPSJ,mceuen:1999PRL_NT_Backscattering}.
The corresponding wavefunction
is given by
\begin{equation}
\psi^0_{s,\bk}(\br)=\frac{1}{\sqrt{2}}\left( \begin{array}{c}
1\\
s\,e^{i\theta_\bk}\end{array}\right)e^{i\bk\cdot\br}\,,
\label{eq:solH0}
\end{equation}
where $\theta_\bk$ is the
angle of the wavevector $\bk$ with respect to the $x$-axis.
Equation~(\ref{eq:solH0}) may be viewed as having the pseudospin vector
being parallel and anti-parallel to the wavevector $\bk$ for electronic states
in the upper ($s=1$) and the lower ($s=-1$) band, respectively 
(Fig.~1B)~\cite{Ando:JPSJ,mceuen:1999PRL_NT_Backscattering}.
As the spin plays a role in the dynamics
of neutrinos, the present pseudospin is similarly important in
the quasiparticle dynamics of graphene~\cite{Ando:JPSJ,
mceuen:1999PRL_NT_Backscattering}.

Now, let us consider a 1D external periodic potential
$V(x)$ applied to graphene.
The potential is taken to vary much more slowly than the carbon-carbon distance
so that inter-valley scattering can be
neglected~\cite{Ando:JPSJ,mceuen:1999PRL_NT_Backscattering}.
Under this condition and for low-energy quasiparticle states
whose wavevectors are close to the $\bK$ point,
the Hamiltonian reads
\begin{equation}
H=\hbar v_0\left(
-i\sigma_x\partial_x+\sigma_y k_y+
{I}\ V(x)/\hbar v_0\right),
\label{eq:newH}
\end{equation}
where $I$ is the $2\times2$ identity matrix.
The eigenstates and eigenenergies of the Hamiltonian $H$ in Eq.~(\ref{eq:newH}) 
may be obtained numerically in the general case or analytically 
for small $\bk$.

It has been predicted that, in a graphene superlattice with
a slowly varying 1D periodic potential or a 2D 
periodic potential of rectangular symmetry, 
the group velocity of its low-energy charge carriers
is renormalized anisotropically~\cite{park:2008NatPhys_GSL}.
Unlike bare graphene which has an isotropic (zero mass)
relativistic energy dispersion (Figs.~1B, 1C and 1D),
graphene under some specific superlattice potentials
displays extremely anisotropic quasiparticle energy dispersion:
the group velocity near the Dirac point along 
the direction perpendicular to the periodicity of the potential
vanishes while the one parallel to the periodicity direction is
intact~\cite{park:2008NatPhys_GSL}.

We consider a Kronig-Penney type of potential
with barrier height $U_0$, lattice period $L$, and barrier width $w$,
periodic along the $x$ direction (Fig.~1A).
These potential parameters can be tuned so that the group velocity 
of the quasiparticles (with wavevector close to the Dirac point)
along the $y$ direction vanishes~\cite{park:2008NatPhys_GSL}.
We shall focus on a graphene superlattice
under one of these conditions ($U_0=0.72$~eV, $L=10$~nm, and $w=5$~nm)
\cite{park:2008NatPhys_GSL}. The parameters used here are experimentally
feasible as shown in recent studies~\cite{meyer:123110,
marchini:2007PRB_Graphene_Ru,vazquez:2008PRL_Graphene_SL,pan:2007condmat_Graphene_SL}.
Later, we will relax the special condition
to confirm the robustness of the predicted supercollimation.

The quasiparticle energy dispersion
of this superlattice (Figs.~1E, 1F and 1G)
shows that, not only the group velocity
of quasiparticles at the Dirac point along the $y$ direction vanishes,
there is hardly any dispersion along the $k_y$ direction within
a good fraction of the supercell Brillouin zone (Figs.~1F and 1G). 
This portion of the energy dispersion in this superlattice thus
is well described by the relation
\begin{equation}
E_s(\bk)=s \hbar v_0 |k_x|\,.
\label{eq:E_kx}
\end{equation}
The deviation of the actual energy dispersion from that of Eq.~(\ref{eq:E_kx})
is less than 5\% for $\bk$ vector as large as 40\% of the supercell
Brillouin zone considered (Figs.~1F and 1G).
This is an equation for wedges.
Thus, for some specific superlattice potentials, graphene turns
from a zero-gap semiconductor into
a quasi-1D metal with a finite and constant density
of states about the Dirac point energy.
We shall call this class of graphene superlattices 
as special graphene superlattices (SGSs).

\begin{figure}
\includegraphics[width=1.0\columnwidth]{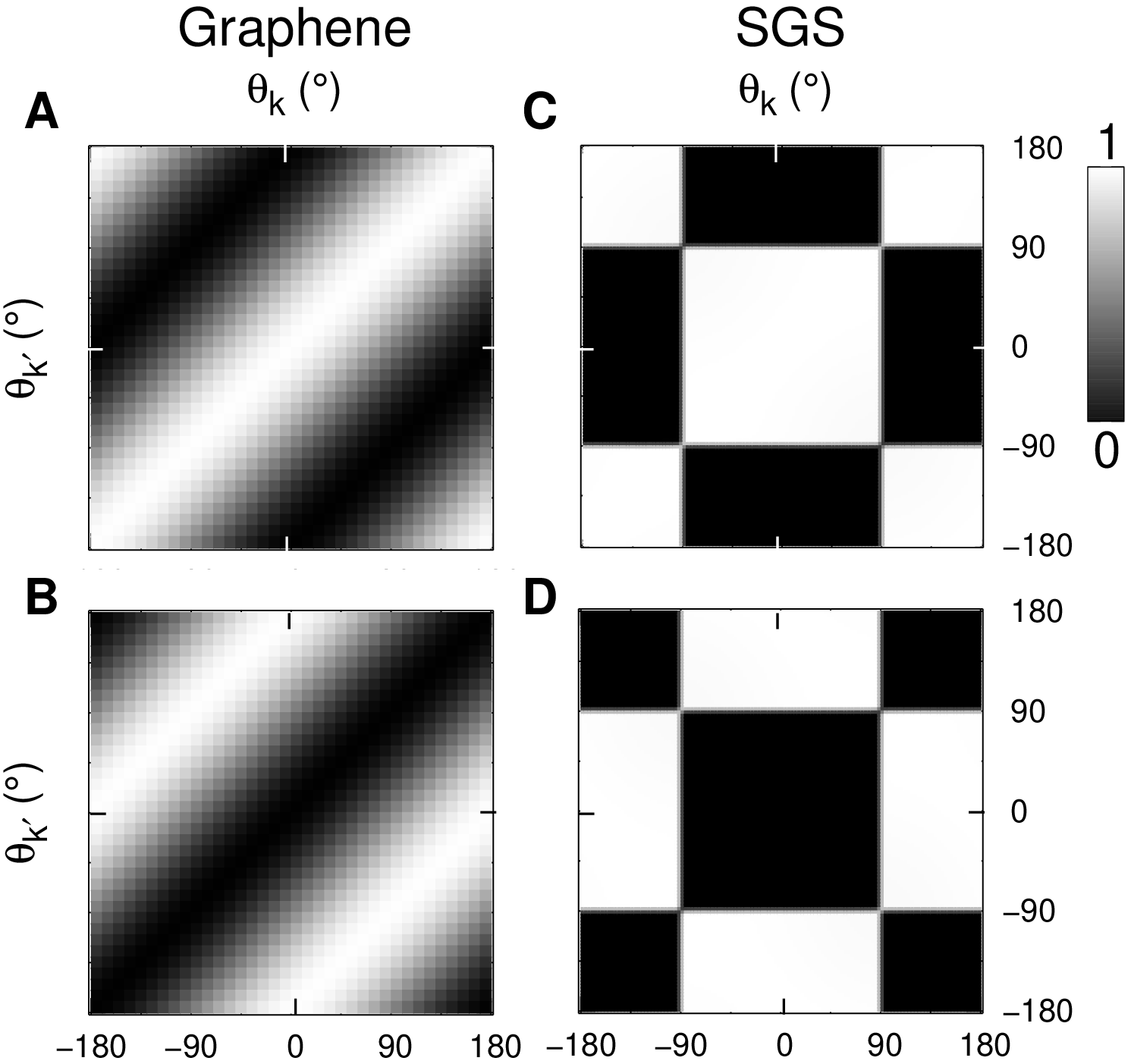}
\caption{{Pseudospin collapse in a special graphene superlattice.}
({\bf A}) and ({\bf B}) Calculated overlap
of two quasiparticle states $\psi^0_{s,\bk}(\br)$ and $\psi^0_{s',\bk'}(\br)$,
$\left|\left<\psi^0_{s',\bk'}|e^{i(\bk'-\bk)\cdot\br}|\psi^0_{s,\bk}\right>\right|^2$,
in graphene versus $\theta_\bk$ and $\theta_{\bk'}$ which are the angles between
the $x$ axis and wavevectors $\bk$ and $\bk'$ ($|\bk|=|\bk'|=0.1\pi/L$), respectively.
The overlap is shown in a gray scale (0 in black and 1 in white).
The two states are in the same band ($s'=s$) in {\bf A}
and are in different bands ($s'=-s$) in {\bf B}.
({\bf C}) and ({\bf D}) Same quantities as in {\bf A} and  {\bf B} for
the considered SGS ($U_0=0.72$~eV, $L=10$~nm and $w=$~5nm).}
\label{Fig2}
\end{figure}

Along with the quasiparticle energy dispersion, 
the internal pseudospin symmetry of the electronic states in SGSs also undergoes a dramatic
alteration. We calculate numerically the overlap
$\left|\left<\psi_{s',\bk'}|e^{i(\bk'-\bk)\cdot\br}|\psi_{s,\bk}\right>\right|^2$
of two quasiparticle states $\psi_{s,\bk}(\br)$ and $\psi_{s',\bk'}(\br)$.
If there were no external periodic potential, this overlap
would be simply $(1+ss'\  \cos\ (\theta_{\bk'}-\theta_\bk))/2$ as seen from Eq.~(\ref{eq:solH0})
(Figs.~2A and 2B).
The same overlap in the SGS (Figs.~2C and 2D)
is however dramatically different from that in graphene,
and can be well described by
$
\left|\left<\psi_{s',\bk'}|e^{i(\bk'-\bk)\cdot\br}|\psi_{s,\bk}\right>\right|^2=
(1+ss'\ {\rm sgn}\ (k_x)\ {\rm sgn}\ (k_x'))/2\,.
$
This behavior is robust for a wide range of the magnitudes of $\bk$ and $\bk'$,
extending over a good fraction of the supercell Brillouin zone
with a high degree of accuracy.
The eigenfunctions of an SGS, for states with small $\bk$,
can be deduced from this result, together with results on
the numerically obtained wavefunctions, (also from analytic
calculation: see Supporting Information)
as having the form of
\begin{equation}
\psi_{s,\bk}(\br)=e^{if(x)}\frac{1}{\sqrt{2}}\left( \begin{array}{c}
1\\
s\ {\rm sgn}\ (k_x)\end{array}\right)e^{i\bk\cdot\br},
\label{eq:solH}
\end{equation}
where $f(x)$ is a real function.
Thus, the spinor in Eq.~(\ref{eq:solH}) is an eigenstate of $\sigma_x$.
Therefore, the direction of the pseudospin
is quantized so that it is either parallel or anti-parallel to the $x$ direction,
which is the direction of the periodicity of the superlattice potential (Fig.~1E),
and not to the wavevector $\bk$ as is the case in pristine graphene (Fig.~1B).
In other words, the pseudospin in the SGS collapses into a backward (`left')
or a forward (`right') state. The resulting quasi-one-dimensionality in the
energy dispersion relation and in the pseudospin of quasiparticles in the SGS
significantly changes the already unique properties of graphene.

\begin{figure*}
\includegraphics[width=1.6\columnwidth]{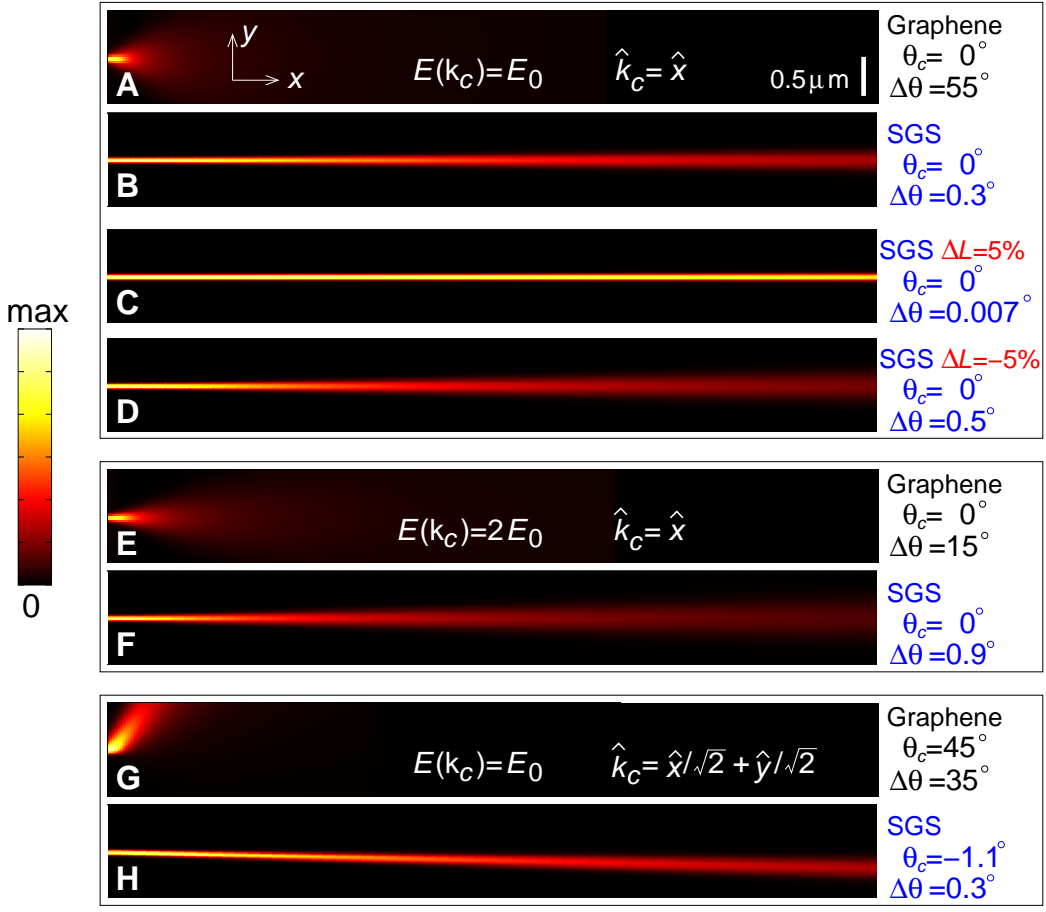}
\caption{{Special graphene superlattice as an electron supercollimator.}
({\bf A}), ({\bf E}) and ({\bf G}) Time-integrated probability density of an electron
wave packet, $\int_0^\infty|\Psi(x,y,t)|^2dt$, in graphene.
The initial ($t=0$) wave packet is a Gaussian localized at the coordinates origin
(middle of the left edge of each panel)
$|\Psi(x,y,0)|^2\sim \exp\left[-\left({x^2}/{2\sigma_x^2}+{y^2}/{2\sigma_y^2}\right)\right]$
where $2\sigma_x=200$~nm and $2\sigma_y=40$~nm.
The wave packet in wavevector space is set to be localized around a specific $\bk_c$.
In {\bf A}, $\bk_c$ is set by
$E(\bk_c)=E_0=\hbar v_0\ 0.1\pi/L=0.02$~eV
and $\hat{k_c}=\hat{x}$. 
In {\bf E}, $\bk_c$ is set by $E(\bk_c)=2E_0$ and $\hat{k_c}=\hat{x}$.
In {\bf G}, $\bk_c$ is set by  $E(\bk_c)=E_0$ and $\hat{k_c}=\hat{x}/\sqrt{2}+\hat{y}/\sqrt{2}$.
$\theta_c$ denotes the angle (defined with respect to the $x$-direction) along
which direction the intensity is maximal
and $\Delta\theta$ denotes the angular spread which gives half the maximum intensity when
the angle is at $\theta_c\pm\Delta\theta$.
({\bf B}), ({\bf F}) and ({\bf H}) Same quantities as in {\bf A}, {\bf E} and {\bf G} for
the considered SGS ($U_0=0.72$~eV, $L=10$~nm and $w=5$~nm), respectively.
({\bf C}) and ({\bf D}), Same quantities as in {\bf B} for graphene superlattices
corresponding to a superlattice potential that is otherwise the same as the SGS studied but
with a period $L$ change of $\Delta L/L=5$\% and $\Delta L/L=-5$\%, respectively.}
\label{Fig3}
\end{figure*}

The quasi-one-dimensionality and specific chiral nature of the SGS
makes it a natural candidate for electron supercollimation.
It is indeed the case that, when a wavepacket of electron is injected into an SGS, 
the propagating packet exhibits essentially no spatial spreading, i.\,e.\,,
electron beam supercollimation is realized (Fig.~3). 
We have calculated the time-evolution of a gaussian wave packet
(with spatial extent along the $x$ and the $y$ direction given by
$2\sigma_x=40$~nm and $2\sigma_y=200$~nm, respectively)
composed of states in the first band above the Dirac point 
energy with a central wavevector $\bk_c$ (Fig.~3).
To provide a measure of the electron beam collimation, we compute the
angle $\theta_c$ in which direction the beam intensity is maximum
and the angular spread $\Delta\theta$ which gives half the maximum intensity when
the angle is at $\theta_c\pm\Delta\theta$.
For a central wavevector $\bk_c$ parallel to the $x$ direction with energy
$E(\bk_c)=E_0=\hbar v_0\ 0.1\pi/L=0.02$~eV, 
the angular spread $\Delta\theta$ in pristine graphene is
$\Delta\theta=55~^\circ$ (Fig.~3A), whereas
in the SGS, $\Delta\theta=0.3~^\circ$ (Fig.~3B),
about 200 times smaller than in graphene.
Specifically, in the SGS, the spread of the wave packet
in the $y$ direction after proceeding 0.1~mm along the $x$ direction 
is only $500$~nm.
Therefore, supercollimation of currents of nanoscale 
width in the SGS can, in principle, be achieved and maintained
as long as the ballistic transport occurs in the system.

Even when the experimental situation deviates from the ideal conditions
for SGSs, supercollimation persists. Hence the phenomenon is quite robust.
First, for example, if we consider
an imperfection in making a superlattice potential
such that the periodicity is slightly larger or smaller ($\Delta L/L=\pm5$\%),
the calculated time-evolution
of a gaussian wave packet shows the angular spread $\Delta\theta=0.007^\circ$
and $0.5^\circ$, respectively (Figs.~3C and 3D).
Second, when considering doped SGSs or high energy electron injection such that 
$E(\bk_c)$ is doubled, the angular spread is still very small (Fig.~3F).
Third, even when $\bk_c$ is off from
the collimation direction ($+x$)
by $45~^\circ$, the angular deviation is still negligible ($\theta_c=-1.1~^\circ$) (Fig.~3H).
In graphene, on the other hand,
the propagation direction and spread of the wave packet sensitively depends
on the magnitude and the direction of the central
wavevector $\bk_c$ (Figs.~3E and 3G).
This robustness here is quite contrary to the case in optics where
the efficiency of supercollimator, superprism
or superlens~\cite{pendry:sciam}
depends sensitively on the magnitude and the direction of the light wavevector
and in general provides a very narrow effective
bandwidth~\cite{kosaka:1212,wu:2003JLT,rakich:2006nmat,joannopoulus:book}.
From our calculations, we expect that the predicted supercollimation 
be observable in SGSs over a wide operation range.

\begin{figure}
\includegraphics[width=1.0\columnwidth]{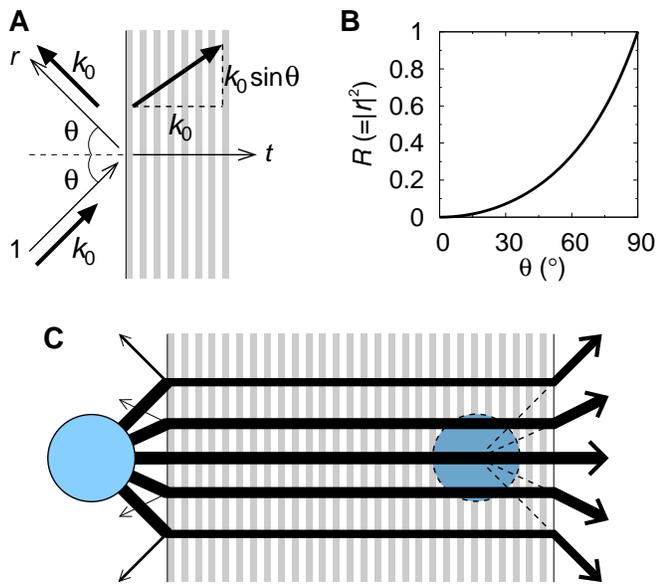}
\caption{{Reflection and transmission at a graphene $-$ special graphene superlattice interface and virtual imaging.}
({\bf A}) Schematic diagram showing the incident, the reflected and the
transmitted wave (the band index is set to $s=1$) at a graphene $-$ SGS interface,
with the relative amplitudes being 1, $r$ and $t$, respectively.
Thick arrows represent the wavevectors of the corresponding waves.
The incidence and reflection angle is $\theta$.
({\bf B}) Reflectance $R=|r|^2$ versus the incidence angle $\theta$.
({\bf C}) Schematic diagram showing the propagation of electron waves
in graphene $-$ SGS $-$ graphene geometry.
Thickness of each arrow is proportional to the actual intensity of the wave.
A virtual image (dashed disk) is formed
at a place far from the actual wave source (solid disk).}
\label{Fig4}
\end{figure}

Lastly, we consider the tunneling properties of injected electrons 
into SGSs from pristine graphene, which provides another measure of
the efficiency of electronic devices based on SGSs.
When an electron is injected into an SGS with an incidence angle
$\theta$ from the graphene side, the wavevector of incident 
electron is given by 
$\bk_i=k_0\ \cos\theta\ \hat{x}+k_0\ \sin\theta\ \hat{y}$ and 
those of the reflected and transmitted electrons by
$\bk_r=-k_0\ \cos\theta\ \hat{x}+k_0\ \sin\theta\ \hat{y}$
and $\bk_t=k_0\ \hat{x}+k_0\ \sin\theta\ \hat{y}$,
respectively (Fig.~4A).
Here, we have made use of the continuity of the transverse component of the wavevector
and conservation of energy, together with the novel dispersion relation given by
Eq.~(\ref{eq:E_kx}). 
Using the continuity of the wavefunction in the system described by Eqs.~(\ref{eq:solH0})
and (\ref{eq:solH}),
we find that the reflectance $R=|r|^2$ is
\begin{equation}
R(\theta)=\tan^2\frac{\theta}{2}\ .
\label{eq:R}
\end{equation}
Interestingly, the reflectance is independent of the specific
form of the external periodic potential of the SGS.
Equation~(\ref{eq:R}) indicates that the transmittance is large for most incidence angles.
For example, even at $\theta=45~^\circ$, the reflectance is less than 20~\%
(Fig.~4B).
Therefore, the SGS is not only an excellent electron supercollimator
but also a good transmitter in a graphene-SGS-graphene junction (Fig.~4C).
Utilizing this property, an immediate application
could be made to demonstrate an electronic analogue of virtual imaging 
in this configuration (Fig.~4C).

Given the recent rapid progress in
graphene superlattices fabrication~\cite{meyer:123110,
marchini:2007PRB_Graphene_Ru,vazquez:2008PRL_Graphene_SL,pan:2007condmat_Graphene_SL}, 
the manipulation of electrons in ways similar to that of photons
in optics by using the supercollimation effect discussed here together with other
optics analogues~\cite{Altshuler:Sci,katsnelson:2006NatPhys_Graphene_Klein}
is expected to soon be practicable. The SGSs have the promise of playing a unique
role in devices based on the synergetic importing of concepts and techniques well
developed in optics to electronics.

{\bf Acknowledgement.} We thank D. S. Novikov and J. D. Sau
for fruitful discussions.
This research was supported by NSF grant DMR07-05941 and by DOE grant DE-AC02-05CH11231.
Y.-W.S. was supported by KOSEF grant R01-2007-000-10654-0 and by
Nano R\&D program 2008-03670 through the KOSEF funded by the Korean
government (MEST).
Computational resources have been provided by NPACI and NERSC.

\newpage
\section*{Supporting Online Material: Analytical solution of electronic states of special graphene superlattices}
When graphene is in a 1D external periodic potential
$V(x)$,  
the Hamiltonian for electrons whose wavevectors 
are close to the $\bK$ pointreads
\begin{equation}
H=\hbar v_0\left(
-i\sigma_x\partial_x+\sigma_y k_y+
{I}\ V(x)/\hbar v_0\right),
\label{eq:newH2}
\end{equation}
where $I$ is the $2\times2$ identity matrix.
In this case, we assume that 
the potential varies much more slowly than the carbon-carbon distance.
So, the inter-valley scattering can be
neglected safely~\cite{ando:1998JPSJ_NT_Backscattering,mceuen:1999PRL_NT_Backscattering}.

If a transform $H'=U^\dagger H U$ is applied to Eq.~(\ref{eq:newH2}) with the unitary matrix
$U=\exp\left(
-i\sigma_x\ \alpha(x)/2\right)$,
where $\alpha(x) = 2\int_0^xV(x')dx'/\hbar v_0$
(we assume that a constant is subtracted from $V(x)$ to set its average to zero),
the resulting Hamiltonian $H'$ reads
\begin{equation}
H'=\hbar v_0\left(
-i\sigma_x\partial_x+\left(\ \cos\alpha(x)\ \sigma_y -
\sin\alpha(x)\ \sigma_z
\ \right)\ k_y
\right).
\label{eq:UnewHU}
\end{equation}
(A similar transform was applied to the Hamiltonian of
a carbon nanotube under a sinusoidal
potential for the specific case of finding the band gap opening behavior
at the supercell Brillouin zone
boundary~\cite{PhysRevLett.87.276802,novikov:235428}.)
The terms having $k_y$ could be treated as a perturbation if $k_y$ is small.
The eigenstate of the unperturbed Hamiltonian is given by
\begin{equation}
\psi'_{0\,s,\bk}(\br)=\frac{1}{\sqrt{2}}\left( \begin{array}{c}
1\\
s\ {\rm sgn}\ (k_x)
\end{array}\right) e^{i\bk\cdot\br}.
\label{eq:solUHnewU}
\end{equation}
Within second order
perturbation theory, using Eqs.~(\ref{eq:UnewHU}) and (\ref{eq:solUHnewU}),
the energy eigenvalue of the superlattice is
\begin{widetext}
\begin{equation}
E_{s,\bk}=s\ \hbar v_0 |k_x|
+
\hbar v_0 k_y^2 \sum_{s',\bG}
\frac{\left|\left<\psi'_{0\,s',\bk+\bG}\left|
\cos\alpha(x)\ \sigma_y - \sin\alpha(x)\ \sigma_z
\right|\psi'_{0\,s,\bk}\right>\right|^2}{s|k_x|-s'|k_x+G_x|}\,,
\label{eq:E_pert}
\end{equation}
\end{widetext}
and the wavefunction is
\begin{widetext}
\begin{equation}
\psi'_{s,\bk}(\br)=\psi'_{0\,s,\bk}(\br)
+
k_y \sum_{s',\bG}
\frac{\left<\psi'_{0\,s',\bk+\bG}\left|
\cos\alpha(x)\ \sigma_y - \sin\alpha(x)\ \sigma_z
\right|\psi'_{0\,s,\bk}\right>}{s|k_x|-s'|k_x+G_x|}
\,\psi'_{0\,s,\bk+\bG}(\br)
\,.
\label{eq:psi_pert}
\end{equation}
\end{widetext}
Here, $\bG$'s are the superlattice reciprocal lattice vectors
$\bG=m\,\bG_0$, where $\bG_0=(2\pi/L,0)$ and $m$ is an integer.
Therefore, in order for the superlattice to be an SGS, in which
there is negligible dispersion with respect to $k_y$, the
energy shift arising from the perturbation (or $k_y$) should be negligible.
Assuming that $\bk$ is small, the dominant summand is the case when $s'=-s$ and $m=0$.
Thus, if
\begin{equation}
\left<\psi'_{-s,\bk}\left|
\cos\alpha(x)\ \sigma_y - \sin\alpha(x)\ \sigma_z
\right|\psi'_{s,\bk}\right>=
0\,,
\label{eq:critical_cond}
\end{equation}
the superlattice would be an SGS.
Under this condition, the relative deviation of the energy dispersion relation, Eq.~(\ref{eq:E_pert}),
from $E_s(\bk)=s \hbar v_0 |k_x|$ (Eq. (3) in the paper) is $O\left(k_y^2/k_xG_x\right)$
and that of the wavefunction, Eq.~(\ref{eq:psi_pert}), from Eq.~(\ref{eq:solUHnewU})
is $O\left(k_y/G_x\right)$.
Similar quantities in a normal graphene superlattice are
$O\left(k_y^2/k_x^2\right)$ and $O\left(k_y/k_x\right)$, respectively.
The eigenfunction $\psi_{s,\bk}(\br)$ of the Hamiltonian $H$ in Eq.~(\ref{eq:newH2}) is,
to a good approximation,
obtained by $\psi_{s,\bk}(\br)=U\psi_{0\,s,\bk}'(\br)$ and is still
an eigenstate of $\sigma_x$ because $U$ commutes with $\sigma_x$.


\begin{thebibliography}{10}

\bibitem{Houten:EPL}
H. van Houten, B.~J. van Wees, J.~E. Mooij, C.~W.~J. Beenakker, J.~G.
  Williamson, and C.~T. Foxon, Europhys. Lett. {\bf 5},  721  (1988).

\bibitem{Spector:APL}
J. Spector, H.~L. Stormer, K.~W. Baldwin, L.~N. Pfeiffer, and K.~W. West, Appl.
  Phys. Lett. {\bf 56},  1290  (1990).

\bibitem{Sivan:PRB}
U. Sivan, M. Haiblum, C.~P. Umbach, and H. Shtrikman, Phys. Rev. B {\bf 41},
  R7937  (1990).

\bibitem{Molenkamp:PRB}
L.~W. Molenkamp, A.~A.~M. Staring, C.~W.~J. Beenakker, R. Eppenga, C.~E.
  Timmering, J.~G. Williamson, C.~J. P.~M. Harmans, and C.~T. Foxon, Phys. Rev.
  B {\bf 41},  R1274  (1990).

\bibitem{Yacoby:PRL}
A. Yacoby, M. Heiblum, V. Umansky, H. Shtrikman, and D. Mahalu, Phys. Rev.
  Lett. {\bf 73},  3149  (1994).

\bibitem{Buks:Nat}
E. Buks, R. Shuster, M. Heiblum, D. Mahalu, and V. Umansky, Nature {\bf 391},
  871  (1998).

\bibitem{Yang:Nat}
Y. Ji, Y. Chung, D. Sprinzak, M. Heiblum, D. Mahalu, and H. Shtrikman, Nature
  {\bf 422},  415  (2003).

\bibitem{Chang:NatPhys}
D.-I. Chang, G.~L. Khym, K. Kang, Y. Chung, H.-J. Lee, M. Seo, M. Heiblum, D.
  Mahalu, and V. Umansky, Nature Phys. {\bf 4},  205  (2008).

\bibitem{Velsen:PRL}
C.~W.~J. Beenakker, C. Emary, M. Kindermann, and J.~L. van Velsen, Phys. Rev.
  Lett. {\bf 91},  147901  (2004).

\bibitem{Palevski:PRL}
A. Palevski, M. Heiblum, C.~P. Umbach, C.~M. Knoedler, A. Broers, and R.~H.
  Koch, Phys. Rev. Lett. {\bf 62},  1776  (1989).

\bibitem{kosaka:1212}
H. Kosaka, T. Kawashima, A. Tomita, M. Notomi, T. Tamamura, T. Sato, and S.
  Kawakami, Appl. Phys. Lett. {\bf 74},  1212  (1999).

\bibitem{wu:2003JLT}
L. Wu, M. Mazilu, and T.~F. Krauss, J. Lightwave Technol. {\bf 21},  561
  (2003).

\bibitem{Prange:book}
{\em The Quantum Hall Effect}, edited by R.~E. Prange and S.~M. Girvin
  (Springer, New York, 1987).

\bibitem{rakich:2006nmat}
P.~T. Rakich, M.~S. Dahlem, S. Tandon, M. Ibanescu, M. Soljacic, G.~S. Petrich,
  J.~D. Joannopoulos, L.~A. Kolodziejski, and E.~P. Ippen, Nature Mat. {\bf 5},
   93  (2006).

\bibitem{joannopoulus:book}
J.~D. Joannopoulos, S.~G. Johnson, J.~N. Winn, and R.~D. Meade, {\em Photonic
  Crystals: Molding the Flow of Light} (Princeton University Press, Princeton,
  New Jersey, USA, 2008).

\bibitem{novoselov:2005Nat_Graphene_QHE}
K.~S. Novoselov, A.~K. Geim, S.~V. Morozov, D. Jiang, M.~I. Katsnelson, I.~V.
  Grigorieva, S.~V. Dubonos, and A.~A. Firsov, Nature {\bf 438},  197  (2005).

\bibitem{zhang:2005Nat_Graphene_QHE}
Y. Zhang, J.~W. Tan, H.~L. Stormer, and P. Kim, Nature {\bf 438},  201  (2005).

\bibitem{berger:2006Graphene_epitaxial}
C. Berger, Z. Song, X. Li, X. Wu, N. Brown, P.~N. First, and W.~A. de~Heer,
  Science {\bf 312},  1191  (2006).

\bibitem{Shedin:NatMat}
F. Shedin, A.~K. Geim, S.~V. Morozov, E.~W. Hill, P. Blake, M.~I. Katsnelson,
  and K.~S. Novoselov, Nat. Mater. {\bf 6},  652  (2007).

\bibitem{Bolotin:SSC}
K.~I. Bolotin, K.~J. Sikes, Z. Jiang, M. Klima, G. Fudenberg, J. Hone, P. Kim,
  and H.~L. Stormer, Solid State Commun. {\bf 146},  351  (2008).

\bibitem{katsnelson:2006NatPhys_Graphene_Klein}
M.~I. Katsnelson, K.~S. Novoselov, and A.~K. Geim, Nature Phys. {\bf 2},  620
  (2006).

\bibitem{Altshuler:Sci}
V.~V. Chelanov, V. Fal'ko, and B.~L. Altshuler, Science {\bf 315},  1252
  (2007).

\bibitem{Ando:JPSJ}
T. Ando, J. Phys. Soc. Jpn. {\bf 74},  777  (2005).

\bibitem{mceuen:1999PRL_NT_Backscattering}
P.~L. McEuen, M. Bockrath, D.~H. Cobden, Y.-G. Yoon, and S.~G. Louie, Phys.
  Rev. Lett. {\bf 83},  5098  (1999).

\bibitem{park:2008NatPhys_GSL}
C.-H. Park, L. Yang, Y.-W. Son, M.~L. Cohen, and S.~G. Louie, Nature Phys. {\bf
  4},  213  (2008).

\bibitem{meyer:123110}
J.~C. Meyer, C.~O. Girit, M.~F. Crommie, and A. Zettl, Appl. Phys. Lett. {\bf
  92},  123110  (2008).

\bibitem{marchini:2007PRB_Graphene_Ru}
S. Marchini, S. G\"{u}nther, and J. Wintterlin, Phys. Rev. B {\bf 76},  075429
  (2007).

\bibitem{vazquez:2008PRL_Graphene_SL}
A.~L. Vazquez~de Parga, F. Calleja, B. Borca, M.~C. G.~P. Jr, J.~J. Hinarejo,
  F. Guinea, and R. Miranda, Phys. Rev. Lett. {\bf 100},  056807  (2008).

\bibitem{pan:2007condmat_Graphene_SL}
Y. Pan, N. Jiang, J. Sun, D. Shi, S. Du, F. Liu, and H.-J. Gao, \rm
  {http://arxiv.org/abs/0709.2858}  (2007).

\bibitem{pendry:sciam}
J.~B. Pendry and D.~R. Smith, Sci. Am. {\bf 295},  60  (2006).

\bibitem{ando:1998JPSJ_NT_Backscattering}
T. Ando and T. Nakanishi, J. Phys. Soc. Jpn. {\bf 67},  1704  (1998).

\bibitem{PhysRevLett.87.276802}
Talyanskii, V. I., Novikov, D. S., Simons, B. D., and L.~S. Levitov, Phys. Rev.
  Lett. {\bf 87},  276802  (2001).

\bibitem{novikov:235428}
D.~S. Novikov, Phys. Rev. B {\bf 72},  235428  (2005).

\end{thebibliography}
\end{document}